\documentclass[referee]{jfm} 
\usepackage{epsfig}
\usepackage{curves}
\usepackage{natbib}
\title[
Characteristic lengths at moving contact lines ]
{Characteristic lengths at moving contact lines for a perfectly 
wetting fluid: the influence of speed on the dynamic contact angle
           }
  
\author[J. Eggers and H. A. Stone]
{Jens Eggers$^*$ and Howard A. Stone$^\ddagger$}  

\affiliation{  
$^*$Universit\"at Gesamthochschule Essen, Fachbereich Physik,   
45117 Essen, Germany \\  
$^\ddagger$Division of Engineering and Applied Sciences, Harvard University,   
Cambridge, MA 02138, USA    
    }  
  
\pubyear{2002}
\date{?? and in revised form ??}

\begin{document}

\maketitle

\begin{abstract}  
It is common to relate the dynamic contact angle $\theta_d$  
to the relative speed between the substrate and the contact line;  
theory suggests $\theta_d^3 \propto U$. In fact, available physical  
models show that the dynamic angle involves speed logarithmically and  
in a model dependent manner. Experimental data consistent with this  
interpretation is cited.   
\end{abstract}  
 
\section{Introduction}  
 
One area of fluid mechanics that has been the subject of a large
admixture of analysis, experiment and speculation is the subject of
the moving contact line.  A typical situation, common in many coating
processes, refers to the contact line at the intersection of solid,
liquid and gas regions, where the three-phase line moves relative to a
solid substrate.  A basic research question in this subject stems from
the violation of the no-slip condition in the immediate neighborhood
of the three-phase line of contact (e.g. Huh \& Scriven 1971, for reviews
see Dussan V. (1979), de Gennes (1985), and Kistler (1993)). 
As a result, within the usual continuum analysis, the
stress diverges as the contact line is approached and the energy per
unit length of the moving contact line is unbounded. This result may
be viewed as an embarrassment of continuum modeling, but, in fact, it
does indicate the need for a small cut-off length scale in macroscopic
theories, as well as some more input from the physics at smaller
length scales to properly interpret the meaning of any such cut-off
scale.
  
Perhaps the most basic feature of this problem is the aim to relate 
the local dynamic contact angle $\theta_d(x)$, which is the arc tangent
of the slope of the interface at a distance $x$ from the contact line, to the 
local speed $U$ with which the contact line moves over the 
substrate. For the case of a perfectly wetting fluid (vanishing 
equilibrium contact angle $\theta_{eq}=0$) and small $\theta_d$,  one finds
$\theta_d^3 \propto U$, 
which is known as Tanner's law.  Theoretical justification for this 
result has been given (e.g. de Gennes 1985) and various 
generalizations have been offered.  In dimensionless form, the speed 
is reported in terms of the capillary number $ {\cal C}=U\eta/\gamma$, 
which measures the relative importance of 
viscous to surface tension forces,  
where $\eta$ is the fluid viscosity and $\gamma$ the interfacial 
tension.  In fact, the functional form for the contact angle-speed 
relation is commonly written for small angles as $\theta_d^3(x) 
\approx 9 {\cal C} \ln(x/\ell_{micro})$, where $\ell_{micro}$ is 
generally taken as a molecular length (e.g. Leger \&
Joanny 1992).  Typically the 
capillary number varies over many orders of 
magnitude; values $10^{-7} < {\cal C} < 10^{-1}$ are common. The 
prefactor in this formula can be important for interpreting experimental 
data, and so it is reasonable to interrogate more closely the 
functional dependence on speed. 
  
In this communication we wish to comment on one aspect of the moving  
contact line problem that has, perhaps surprisingly, been largely  
neglected and/or unappreciated.  In particular, we note that detailed  
models for the perfectly wetting 
situation actually yield a dynamic contact angle versus speed relation  
\begin{equation}  
\theta^3_d(x) \approx 9{\cal C} \ln \left ( \frac{x}{\ell_{micro}}{\cal  
C}^\beta\right ) ~, 
\label{contactline1}  
\end{equation}where $\beta$ depends on the   
physical model introduced in the neighborhood of the contact line. We 
do not believe that it is necessarily appropriate to simply suppress 
the additional dependence on speed (i.e. ${\cal C}$) by replacing the 
argument of the logarithm by either $\ell_{macro}/\ell_{micro}$, where 
these two lengths scales are taken as constants, or 
$x/\ell_{micro}$. Because of the large variation in $\cal C$, not 
including this additional factor of capillary number when using 
(\ref{contactline1}) to interpret dynamical experiments may lead to 
significant discrepancies between theory and experiment. Here 
we outline the basic idea behind (\ref{contactline1}) and present 
experimental evidence that supports the above interpretation. 

Another point that has received insufficient attention is the range of
validity of equation (\ref{contactline1}). Near the contact line,
(\ref{contactline1}) breaks down where $x$ is of the same order as
$\ell_{micro}$. This restriction is evident as the general structure
comes from a balance of viscous and surface tension forces alone.  Not
surprisingly, we estimate below that the microscopic scale is between
several Angstroms and tens of Angstroms, depending on the microscopic
forces assumed to be acting near the contact line.

Towards large scales, $x$ is commonly taken to be a static scale such
as the capillary length or the size of a spreading drop (de Gennes
1985).  Nevertheless, it should be noted that the flow near a moving
contact line often resembles a coating flow, similar to the classical
problem studied by Landau and Levich (e.g. Levich 1962). This leads to
the appearance of another, {\it dynamical} length scale, that can
become much smaller than the capillary length as the capillary number
is small, which is typically the case. Thus a meaningful comparison
between (\ref{contactline1}) and a macroscopic measurement of the
dynamical contact angle might require a spatial resolution
significantly below 1/10 or even 1/100 of the capillary length.
  
In the next section we will introduce two different models commonly
used to treat moving contact line problems such as a spreading drop 
or a tape plunging into a pool of fluid. Then, in the third section 
we show that the lubrication equations corresponding to both models 
have similarity solutions for the interfacial shape that fix the 
functional dependence on the capillary number. The fourth section 
discusses the dynamical problem that equation (\ref{contactline1}) 
has to be matched to on an appropriate {\it outer} length scale. 
In the fifth section we explain measurable consequences of 
the two models for the dynamic contact angle, and discuss an
experiment that helps to distinguish between them. We close with
a summary and possible directions of future work. 

\section{The model}  
 
The usual dynamic balance for the steady flow ``far'' from the contact
line involves capillary and viscous stresses. As the contact line is
approached, the capillary-viscous flow leads to a stress
singularity. A number of different physical effects have been
suggested to relieve the singularity, and these either account for the
fact that on very small length scales van der Waals forces act to
maintain a finite thickness liquid layer on the solid substrate, or
that at very high shear rates the boundary conditions and the
transport coefficients of the fluid are likely to be altered. Which
model is appropriate might depend on the physical system at hand, or
be a combination of the above. In Table \ref{t1} we provide a short
overview of proposed physical models for flow in the neighborhood of a
contact line; see also McKinley \& Ovryn (1998). Most recently, 
there has been a considerable effort to base the understanding 
of the contact line physics on a microscopic, particle-based 
description, see for example Koplik et al (1989), Ruijter et al. (1999),
and Abraham et al. (2002). The so-called ``diffuse interface model'' 
(see e.g. Seppecher (1996), Chen et al. (2000), and Pomeau (2002)) 
represents an intermediate approach,
which models the liquid-gas interface as a Cahn-Hilliard fluid.
This allows for example the extraction of effective interface equations 
(Pismen \& Pomeau (2000)) {\it different} from those proposed 
by de Gennes's (1985).
 
Below we  restrict our attention to two different models which 
have proved particularly popular. The results are sufficient to highlight 
the measurable differences between different physical mechanisms. In 
model I, due to de Gennes and coworkers (e.g. 
Hervet \& de Gennes 1982, de Gennes 1985), van der Waals forces are 
taken into account, so very close to the contact line there is a 
balance between surface tension and van der Waals stresses alone. In 
model II, proposed for example by Huh and Mason (1977)  and 
Hocking (1977), the fluid is allowed to slip across the solid 
surface over a small slip length. 
 
\begin{table}
  \begin{center} 
    \leavevmode 
\begin{tabular}{ccccccccc} 
mechanism & reference & \\ \hline 
van der Waals & Hervet \& de Gennes (1984) & \\ 
Navier slip & Huh \& Scriven (1971) & \\ 
nonlinear slip &  Thompson \& Troian (1997) & \\ 
shear thinning& Gorodtsov (1990)  &  \\ 
diffuse interface &  Seppecher (1996)  & \\
generalized Navier slip &  Shikmurzaev (1997) & \\
\end{tabular}  
  \end{center} 
\caption{Different models for the flow in the neighborhood of the
contact line, with representative references.  }  \label{t1}
\end{table} 
 
For simplicity, we only consider the case of perfectly wetting fluids,
i.e. of zero equilibrium contact angle. Consistent with the local
balances, the interface near the contact line remains nearly flat and
we can use lubrication theory to describe the fluid motion.  This
approach amounts to a significant simplification of the mathematical
treatment relative to the full two-dimensional flow problem (Cox
1986), but agrees with the full calculation when the dynamic contact
angle is small. There are numerous indications that the small-angle
theory in fact remains valid for slopes of order unity.  For example,
$\theta_d^3/9$ in equation (\ref{contactline1}) differs by only $2\%$
from the full expression (Cox 1986), derived without the benefit of
lubrication theory, up to a slope of 1.  Also, Fermigier \& Jenffer
(1991) reported that small-angle theory holds experimentally up to an
angle of $100^{\circ}$.
  
\begin{figure}
\begin{center}  
\psfig{figure=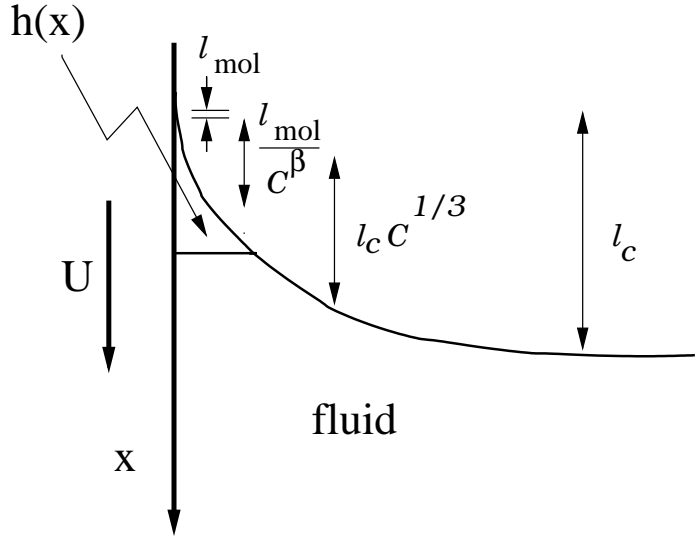}  
\end{center}  
\caption{A typical application involving a moving contact line: a plate
plunges with velocity $U$ into a liquid-filled container; the capillary
length $\ell_c = 
\sqrt{\frac{\gamma}{\rho g}}$.  Since we
assume wetting fluids, the meniscus creeps up the plate opposite the
direction of motion and a nonzero dynamic contact angle $\theta_d (x)$
is established.  In the experiment of Marsh et al. (1993) referred to
below the plate is replaced by a cylinder which can be tilted at
different angles $\alpha$.  We also schematically indicate the
different length scales relevant for this problem.}
\label{exp}  
\end{figure}  
  
To be able to describe an experiment like a flat plate plunging with
velocity $U$ into a reservoir of fluid (see Figure \ref{exp}), it is
necessary to include other terms beyond the lubrication terms close to
the contact line. Namely, we keep the full curvature term and include
gravity. By doing this, the model is able to describe the crossover to
a purely static, horizontal surface far away from the dynamical
region.  A basic unit of length is the capillary length $ \ell_c =
\sqrt{\frac{\gamma}{g \rho}} $, which dictates the scale of the
interface curvature far away from the contact line.  In the van der
Waals model I, which accounts for pressure variations owing to
capillary, van der Waals, and gravitational forces, the equation for
the {\it stationary} profile $h(x)$ (cf Figure \ref{exp}) is
\begin{equation}  
\label{PGG}  
\frac{3 {\cal C}}{h^2} = \kappa' + 3a^2 \frac{h'}{h^4} - \ell_c^{-2} ,   
\end{equation}  
where $\kappa$ is the curvature and a prime refers to differentiation 
with respect to $x$. Note that a positive ${\cal C}$ corresponds to 
the plate plunging {\it into} the fluid. A brief derivation of model 
I, as well as model II below, is given in appendix 1.  The microscopic 
 length parameter $a$, defined by 
\begin{equation}  
\label{a}  
a^2 = \frac{A}{6\pi\gamma}~, 
\end{equation} 
measures the strength of van der Waals forces relative  
to interfacial forces and is typically very small  (on the order 
of Angstroms).  
 
Another distinct approach for the flow near the contact line is 
to introduce  slip at the boundary, consistent with 
allowing the contact line to move parallel to the wall at a finite   
speed; the slip is a function of the shear rate. The simplest such
law, introduced by Navier in the same paper that also enunciated the
Navier-Stokes equation (Navier 1823), is
\begin{equation}  
\label{Navier}  
u|_{y=0}-U = \lambda\frac{\partial u}{\partial y}|_{y=0}
\end{equation}(see also Huh \& Scriven 1971).   
Here $U$ is the speed of the moving boundary, $y=0$ denotes the 
solid-liquid boundary, and $\lambda$ is the so-called slip length. 
A more complicated version of (\ref{Navier}), in which 
$\lambda$ is itself a non-linear function of the shear rate,
has been proposed in Thompson \& Troian (1997).
A standard calculation (appendix 1), leads to the  
analogue of equation (\ref{PGG}) for the slip model II,  
\begin{equation}  
\label{Hocking}  
\frac{3 {\cal C}}{h^2} = \kappa' -  
3\lambda\frac{(1-\kappa')}{h} - \ell_c^{-2} .
\end{equation}  
The slip length $\lambda$ is usually considered to be on the order of 
tens of Angstroms. 

\section{Scaling solutions near the contact line }  
 
We now focus on the immediate neighborhood of the contact line, which we  
assume to be at $x=0$. Owing to the flatness of the interface   
we can assume that $\kappa \approx -h''$ and gravitational
influences can be neglected, but dynamical (viscous)   
effects have to be included.   
In the case of model I, equation (\ref{PGG}) reduces to    
\begin{equation}  
\label{lub1}  
\frac{3 {\cal C}}{h^2} = -h''' + 3a^2 \frac{h'}{h^4}.  
\end{equation}  
To make the dependence on parameters explicit,  
we note that (\ref{lub1}) has the exact   
scaling solution
\begin{equation}  
\label{scal1}  
h(x) = \frac{a}{{\cal C}^{1/3}}\phi_1(a^{-1}{\cal C}^{2/3}x),  
\label{modelone-shape} 
\end{equation}  
where $\phi_1 $ depends on the similarity variable  
$\xi_1 = a^{-1}{\cal C}^{2/3}x$ and satisfies the equation  
\begin{equation}  
\label{sim1}  
\frac{3}{\phi_1^2} = -\phi_1''' + 3\frac{\phi_1'}{\phi_1^4}.  
\end{equation}  
  
Similarly, the lubrication approximation for model II gives  
\begin{equation}  
\label{lub2}  
\frac{3 {\cal C}}{h^2} = -h''' - \frac{3\lambda}{h}h'''~. 
\end{equation}  
In this case the scaling solutions are   
\begin{equation}  
\label{scal2}  
h(x) = \lambda\phi_2({\lambda}^{-1}{\cal C}^{1/3}x),  
\label{modeltwo-shape} 
\end{equation}  
where the similarity variable is now $\xi_2 = 
{\lambda}^{-1}{\cal C}^{1/3}x$ and the similarity 
equation is 
\begin{equation}  
\label{sim2}  
\frac{3}{\phi_2} = -\phi_2'''  
- \frac{3}{\phi_2}\phi_2''' ~. 
\end{equation}  
  
Far away from the contact line in units of the microscopic   
lengths $a$ and $\lambda$, respectively, the solutions should   
be the same, resulting from a balance of classical viscous forces   
and surface tension. Indeed, as $\xi\rightarrow\infty$ 
one finds to leading order   
\begin{equation}  
\label{asymp}  
\phi_{1,2}(\xi) \approx 3^{2/3}\xi\left[\ln(\xi b_{1,2})\right]^{1/3}, \quad
\xi \gg 1~,
\end{equation}  
where the numerical constants $b_{1,2}$ have to be determined by
numerical integration starting from the contact
line. 

The boundary condition at the contact line
incorporates the wetting behavior of the fluid.  The basic assumption
is that there is a microscopic length scale on which static forces
dominate over dynamical ones, and a static profile can be assumed
microscopically close to the contact line.
 
In model I, following Hervet \& de Gennes (1984), we are going to
match $\phi_1(x)$ to a ``maximal'' film solution, corresponding to
very strong wetting, whose thickness only goes to zero at (minus)
infinity.  However this maximal solution very closely approximates
parabolic solutions of (\ref{lub1}) that go to zero at some finite
contact line position (Hervet \& de Gennes 1984).  To leading order,
we take the film solution to be of the form
\begin{equation}  
\label{film}  
\phi_1(\xi_1) = -\frac{1}{\xi_1} + 
\epsilon\exp\left\{\xi_1^3/\sqrt{3}\right\}, 
\end{equation}  
and defer further details to appendix 2. Using (\ref{film}) as  
an initial condition with adjustable parameter $\epsilon$, we 
integrate (\ref{sim1}) towards $\xi_1\rightarrow\infty$. The  
parameter $\epsilon$ is fixed to select the solution with vanishing  
curvature at infinity. Figure \ref{compare} compares this  
solution with the asymptotic form (\ref{asymp}). We plot the  
rescaled slope $\phi_1'$ from the solution of (\ref{sim1}) as  
the full curve and equation (\ref{asymp}) with $b_1 = 1.44$ as the dashed  
curve. This numerical value for $b_1$ differs significantly from  
$b_1 = 0.4\cdot3^{1/6}\approx 0.48 $ given in de Gennes (1985), accounting for 
differences in normalization. We believe the difference is simply 
due to the large values of $\xi$ necessary for integration until 
a true asymptotic value is reached. 
 
Model II, on the other hand, can be extended down to  
$\phi_2(0) = 0$, since the stress singularity was successfully  
removed. A vanishing equilibrium contact angle $\theta_{eq}=0$ can 
thus be implemented by taking the boundary condition  
$\phi_2(0)=\phi_2'(0)=0$ at the contact line and integrating  
(\ref{lub2}) towards $\xi_2\rightarrow\infty$. The corresponding  
value of $b_2$ for the case of the Navier  
slip law was given in Hocking (1992). Thus the two constants, which
establish the form of the interface profile, are  
\begin{equation}  
\label{constants}  
b_1\cong 1.44,\quad b_2 \cong 3^{1/3} \exp(0.74/3)\cong 1.85.
\end{equation}  
  
\section{Crossover to Landau-Levich-type behavior} 
 
We now estimate the range of validity of the solution (\ref{asymp}) 
as one moves farther away from the contact line. These ideas have 
close analogy to the classical analysis of Landau and Levich 
of a dynamical lubrication film (Levich 1962). Evidently, the contact 
line physics plays no role far away from the contact line, so the relevant  
lubrication equation is  
\begin{equation}  
\label{LL}  
\frac{3{\cal C}}{h^2} = -h''' ,
\end{equation}  
which is to be matched to a static profile at large distances. 
This equation has the general solution 
\begin{equation}  
\label{LLshape}  
h(x) = \ell_c{\cal C}^{\alpha_1+1/3}f(x/(\ell_c{\cal C}^{\alpha_1})),  
\end{equation}  
which has to be matched to a static meniscus on the capillary  
scale. This static solution is characterized by a curvature  
that is approximately constant, thus $h''(x)$ must be independent  
of ${\cal C}$. This fact forces $\alpha_1=1/3$, so that we 
have  
\begin{equation}  
\label{LLshape-f}  
h(x) = \ell_c{\cal C}^{2/3}f(x/(\ell_c{\cal C}^{1/3})), 
\end{equation}  
which implies that the crossover will occur on a scale $\ell_c{\cal C}^{1/3}$, 
on which the logarithmic dependence in equation (\ref{asymp}) for the slope  
begins to fail.  
 
\begin{figure}  
\begin{center}  
\psfig{figure=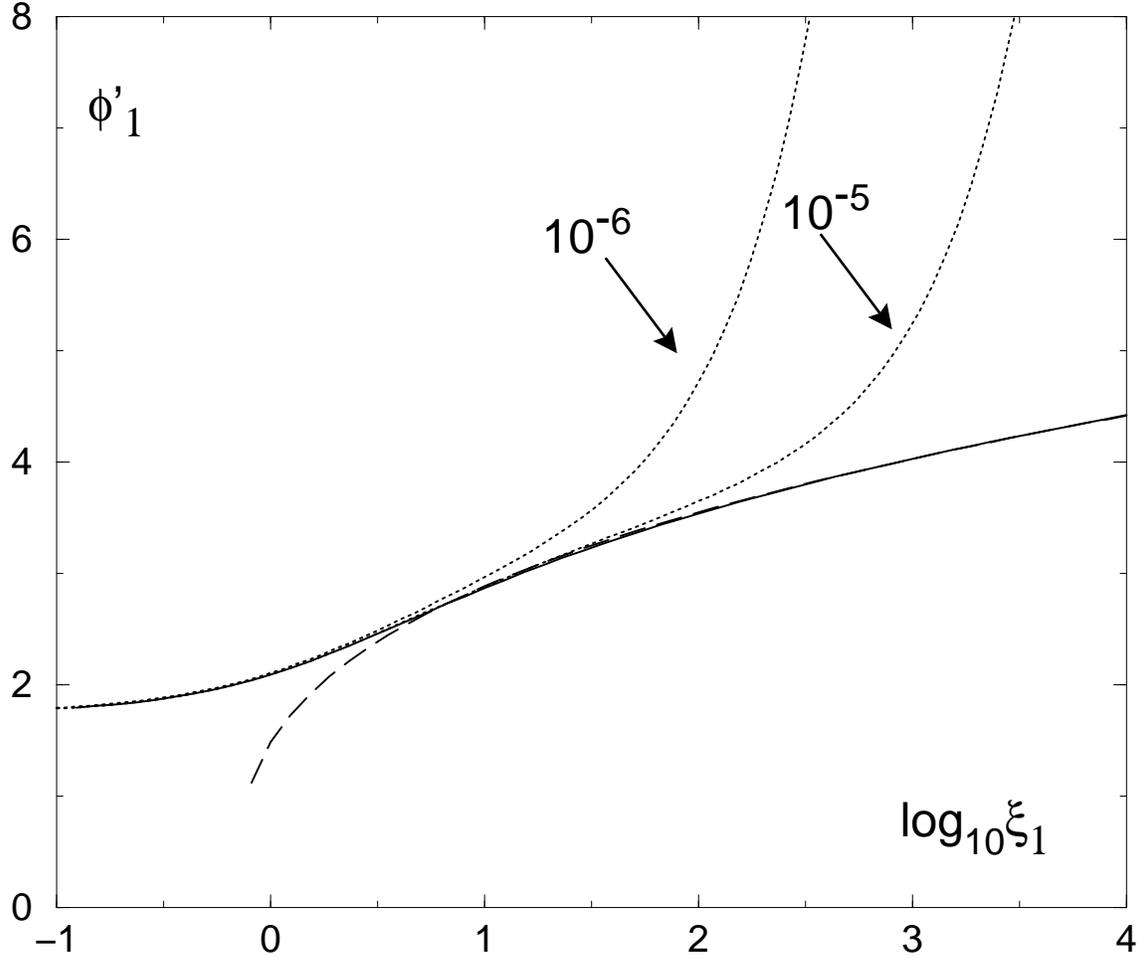} 
\end{center}  
\caption{The rescaled slope $\phi_1' = h'a^{-1}{\cal C}^{1/3}$ versus  
$\log_{10}(\xi_1) = \log_{10}(a^{-1}{\cal C}^{2/3}x)$  
for ${\cal C} = 10^{-5}, 10^{-6}$ in the geometry of Figure \ref{exp}.  
The ratio $a/\ell_c$ equals $10^{-8}$. The full line 
is a solution of (\ref{lub1}), the dashed line corresponds to  
the asymptotic form (\ref{asymp}). The two dotted lines are  
solutions of the full system
(\ref{PGG}) including gravity, marked with their values of
the capillary number.  
Note that for the smaller capillary number the size of the overlap 
region where the asymptotic form (\ref{asymp}) can be applied has 
shrunk to zero. 
   }  
\label{compare}  
\end{figure}

The crossover to scaling of the form of equation (\ref{LLshape-f}) is
demonstrated in Figure \ref{compare}, by showing a full solution of
equation (\ref{PGG}), rescaled according to (\ref{scal1}). Results are
given for two (small) values of the capillary number differing by a
factor of ten.  Again, the free parameter in the maximal film solution
(\ref{film}) is used to shoot for the flat interface corresponding to
the surface of the fluid-filled container. For small values of
$\xi_1$, the solution corresponds to the lubrication form given
before, while on a scale $x /\ell_c\approx {\cal C}^{1/3}$ the
transition to the Landau-Levich region is observed. In rescaled
coordinates $\xi_1$ the location of this crossover should thus be
proportional to ${\cal C}$ itself, as is clearly seen from 
Figure 2. We have chosen the smaller of the two 
values of ${\cal C}$ such that the region
over which the asymptotic form (\ref{asymp}) of the interface can be
applied is zero, to highlight possible problems in comparing
asymptotic solutions with experimental data. To interpret the measured
dynamical contact angle equation (\ref{asymp}) is no longer sufficient, but the
full solution of the similarity equation (\ref{sim1}) has to be
considered.

\section{Comparison with experiment}  
 
It is common practice in the literature (e.g. de Gennes 1985, Cox 1986) to
consider the derivative of the profile $h(x)$, evaluate it at some
macroscopic distance from the contact line $x=\ell_{macro}$, and to
interpret the slope of the interface in terms of the so-called
``dynamical contact angle'', $\tan\theta_d(x)= dh/dx$. This approach is the 
common one taken in experiments as well. Thus, using the
solution (\ref{asymp}) in the similarity forms (\ref{scal1}) and
(\ref{scal2}) for models I and II, respectively, and neglecting
lower-order terms, we obtain
\begin{equation}  
\label{dyn}  
\theta^3_{dyn}(x) = 9{\cal C}\ln(\ell_{macro}/L_{1,2}),  
\end{equation}  
where $L_{1,2}$ are microscopic lengths appropriate for each model.  
There are two fundamental issues with this approach:   
First, depending on the experimental system it is not clear what is the   
best choice for $\ell_{macro}$. Second, what is usually   
taken as a fixed microscopic length $L_{1,2}$ is   
actually strongly dependent on the capillary number.   
Namely, the two models give   
\begin{equation}  
\label{lmicr}  
L_1 = a {\cal C}^{-2/3} / b_1 , \qquad  \hbox{and}\qquad
L_2 = \lambda {\cal C}^{-1/3} / b_2 .  
\end{equation}as established in (\ref{modelone-shape})  
and (\ref{modeltwo-shape}), respectively.  
  
In particular, the ${\cal C}$ dependence that appears in the
microscopic length is different in the two models.  It is also clear
that it is impossible to interpret $L_{1,2}$ directly in terms of some
fixed microscopic length near the contact line, but rather it is a
{\it dynamical} quantity. To our knowledge, this fact has never been
appreciated in either theoretical or experimental work. This
observation appears to be significant, since by comparing the ${\cal
C}$-dependence it potentially allows one to distinguish between
different microscopic models from a macroscopic measurement.  Chen and
Wada (1989) imaged the profile near the contact line of a spreading
droplet and so provided the first experimental confirmation of
(\ref{dyn}). However, owing to the small range of capillary numbers
studied, it is difficult to distinguish between the two lengths
$L_{1,2}$ defined in (\ref{lmicr}).  Below we will therefore
concentrate on another experiment (Marsh, Garoff \& Dussan 1993), which allowed
$\cal{C}$ to be varied over more than two orders of magnitude.
  
It is also interesting to note that the logarithmic dependence   
on capillary number was only obtained in the fully nonlinear treatment   
outlined above. In the classical  studies of the flow in the neighborhood
of the dynamic contact angle, for example in Cox (1986), a matched asymptotic 
analysis is used which further assumes a form involving integer powers 
of ${\cal C}$ and does not recognize that the scale of the inner region
can itself involve ${\cal C}$. 
However the form of the asymptotics (\ref{dyn}), implying a {\it logarithmic}
dependence on ${\cal C}$ in the full solution, clearly shows that 
the profile cannot be expanded in integer powers of {\cal C}. As a result,   
the classical analyses are not able to identify the sort  
of dependencies given by (\ref{lmicr}). Although these dependencies  
are only logarithmic, as mentioned above, 
the capillary number often varies through many  
orders of magnitude in experiments, so the logarithm in (\ref{dyn})  
can in general not be approximated by a constant, as is most often done
(e.g. King 2001).  
  
In a recent experiment, the effect of large variations of ${\cal C}$
on the contact line was investigated very carefully by Marsh et
al. (1993), who measured the dynamic contact angles on a cylinder
plunging at an angle into a liquid bath.  
These authors essentially used the form (\ref{dyn})
to fit the whole shape of the interface close to the contact line, and
included {\it static} contributions to account for the effects of
surface tension and gravity away from the contact line. 
(Note that instead of the third power on the left hand side,  
they actually used a more complicated function $g(x)$, but  
which becomes $g(x)\approx x^3/9$ for small arguments. This limit is   
 relevant  for the small angle case we are studying here.)
  This approach
leaves out dynamical effects of the kind predicted by Landau and
Levich (Levich 1962), which are important on an intermediate scale
between the microscopic ones and the capillary length, and should be
taken into account in a more refined theory.  Marsh et al. (1993)
treat the static contact angle $\theta_{eq}$ (called $\theta_{act}$ 
by the authors) as a free, and possibly ${\cal C}$-dependent, parameter, 
to be determined from experiment. The authors conclude that for
their system  
$\theta_{eq}=0$, which is the case treated here.
  
\begin{figure}
\begin{center} 
\psfig{figure=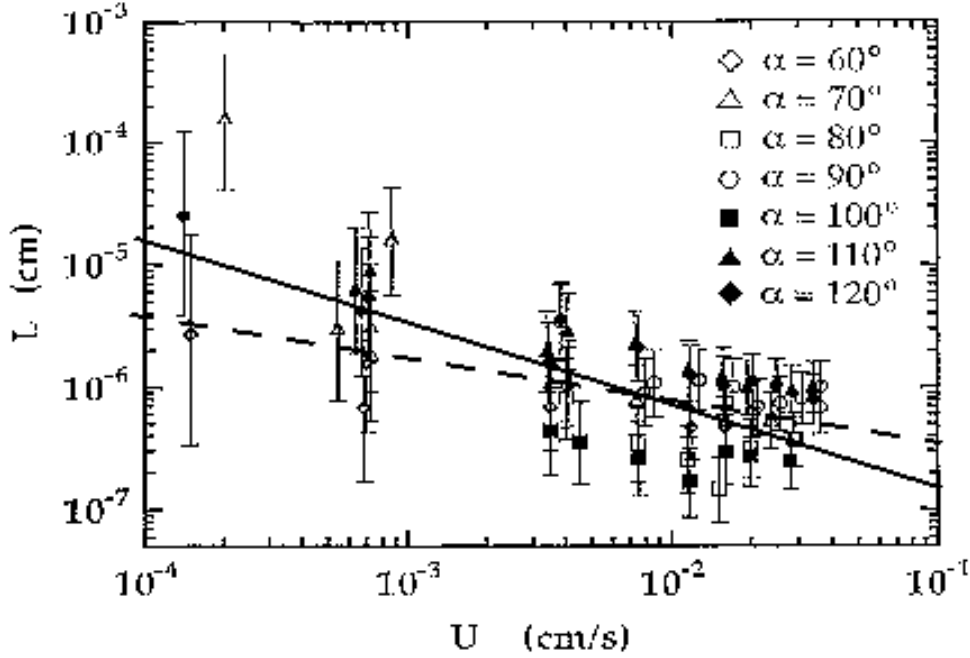,width=5.5in} 
\end{center}
\caption{A plot of the microscopic length $L$, taken from Figure 4 of 
Marsh et al. (1993), as function of the speed $U$. 
Their $L$ is the equivalent of $L_{1,2}$ as given in (\ref{dyn}).
The angle $\alpha$ refers to different tilts of the   
solid relative to the liquid surface. To make a comparison with the 
dynamical length sales $L_{1,2}$ as given in (\ref{lmicr}), we have
added to the figure the solid and the dashed lines with slope 2/3 
and 1/3, respectively.  
  }  \label{MGD}  
\end{figure}  
  
From the fit of (\ref{dyn}) to their data, Marsh et al. (1993) extract
a length $L$, which is found to depend significantly on capillary
number, as suggested by (\ref{lmicr})! They also report $L$ to be
independent of the tilt angle $\alpha$ within experimental error,
which further emphasizes that the response is dominated by local
features.  In Figure \ref{MGD} we present a plot of the measured
length as function of capillary number, and compare it with the slopes
suggested by the van der Waals model I and the Navier slip model II,
respectively.  Although it is difficult to draw firm conclusions owing
to the large scatter in the data, 2/3 seems to be favored. Using the
two different fits plotted in Figure \ref{MGD}, we are also able to
determine the cut-off lengths $a$ and $\lambda$, assuming that the
corresponding physical mechanism is really relevant for the particular
materials involved. We find $a \approx 4$\AA for the van der Waals
model and $\lambda\approx 30$\AA for the slip model. Using the value
of $A = 10^{-20} \hbox{J}$ (Russel, Saville \& Schowalter 1989) for 
the Hamaker constant for
water and an adjacent solid surface, and $\gamma = 0.07 \hbox{N/m}$,
we find $a \approx 1$\AA, consistent with the above value.  However, it is
important to keep in mind that there is no reason why {\it a single}
cutoff mechanism should necessarily dominate in the experiment, which
would lead to still other exponents.  Additional mechanisms for
relieving the contact line singularity are listed in Table \ref{t1};
which is the dominant mechanism could also depend on capillary number
and, in particular, on the type of solid substrate or fluid involved.
 
\section{Conclusions}

We have considered flow local to a moving contact line using a
lubrication approach. Our basic message has been to indicate that the
appearance of logarithmic corrections in capillary number to the usual
``Tanner's law'', $\theta_d^3\propto {\cal C}$, are a general feature
of the mechanical response. The interpretation of the results is that
the ``microscopic'' length scale that is involved when supplying a
small-scale cut-off to relieve the well-known stress singularity in
the moving contact line problem is a dynamical (speed dependent)
quantity. Experimental data consistent with this interpretation is cited,
and microscopic parameters that come from the comparison with 
different theories agree with physical considerations. 

Since important applications of contact line theories apply to angles
up to $180^{\circ}$, it would be very useful to extend the lubrication
theory considered here to a full two-dimensional treatment of the flow
in the corner region. We suspect that large parts of the calculation
in Cox (1986), which erroneously assumes a regular expansion in ${\cal
C}$, could be adapted to a proper similarity description. This means
that the dynamical contact angle has the scaling form $\theta_d =
f({\cal C}, x/L_{1,2})$, where $L_{1,2}$ is one of the {\it dynamical}
length scales defined by (\ref{lmicr}), with corresponding forms for
the velocity field as well.  Such a theory might be able to explain
more recent experiments (Chen, Ram\'e \& Garoff 1995) on moving
contact lines performed at higher capillary numbers, yielding
dynamical contact angles of up to $155^{\circ}$.

Furthermore, the velocity of the contact line
relative to the substrate is in general not {\it perpendicular}
to the contact line, as highlighted 
in recent experiments of droplets running 
down an inclined plane (Podgorski, Flesselles \& Limat 2001). In this case the flow is 
truly three-dimensional, and it may no longer be sufficient to 
simply project the velocity onto the normal to the contact
line (Blake \& Ruschak 1979). Such a three-dimensional description would be 
necessary to complete our understanding of corner singularities 
that form at the back of running drops (Stone et al. 2002), and may apply 
to a range of other contact line phenomena as well. 

\section{Appendix 1: Derivation of lubrication equations}  
 
Here we briefly recall the derivation of the interface, or 
lubrication, equations for thin viscous films (Levich (1962)). For 
pressure-driven flow along the surface and absorbing hydrostatic 
pressure variations into the pressure $p$, the velocity parallel to 
the plate can be represented as a second-order polynomial 
\begin{equation}  
u = a_0 + a_1y + y^2 \frac{p^\prime}{2\eta}~, 
\end{equation}where $y$ 
is the distance normal to the plate. At the free surface $y=h(x)$, 
shear gradients $\partial u/\partial y$ vanish, giving $a_1 = 
-p^\prime h/\eta$.  Finally, from the slip condition (\ref{Navier}) we have 
$a_0=U - \lambda p^\prime h/\eta$. 
 
Since the contact line is stationary, the mass flux through the film 
is zero everywhere, 
$ \int_0^h u(y) ~{\rm d}y = 0 
$, and thus  
\begin{equation}  
\label{zero_flux}  
0 = (U - \lambda p'h/\eta)h - p'h^3/3\eta 
\end{equation} 
is the equation for the film profile $h(x)$. In the presence of van der Waals  
forces, the dynamic pressure in the liquid is  
\begin{equation}  
p = \gamma\kappa - \frac{A}{6\pi h^3} - g\rho x, 
\label{lubrication-pressure}
\end{equation} 
where $A$ is Hamaker's constant. Substituting 
(\ref{lubrication-pressure}) into (\ref{zero_flux}) 
and assuming $\lambda = 0$ gives (\ref{PGG}), while  
$A=0$ at finite $\lambda$ leads to (\ref{Hocking}). 
 
\section{Appendix 2: expansion for the maximal film}  
 
Here we give some more details on the solution of (\ref{lub1}) 
for the ``maximal film''  of Hervet and de Gennes (1984).
The general form of the film  profile is  
\begin{equation}  
\label{phi_0} 
\phi_0 = \frac{1}{\xi}\sum_{i=0}^{\infty} \frac{a_i}{\xi^{6i}}, 
\end{equation} 
where we denote the similarity variable by $\xi$.  
This expansion has {\it no} free parameters, as the values of the  
coefficients $a_i$ are obtained directly from substituting (\ref{phi_0}) 
into (\ref{lub1}). We find 
\begin{equation}  
\label{a_i} 
a_0 = -1~, \quad a_1=-2/5~, \quad a_2=-1764/275, \dots. 
\end{equation} 
 
However, there is a one-parameter family of solutions of  
(\ref{lub1}) that decay for $\xi\rightarrow-\infty$. This
solution is  
found by linearizing around the base solution (\ref{phi_0}),
i.e. $\phi (\xi) = \phi_0 (\xi) + \delta(\xi)$: 
\begin{equation}  
\label{lin} 
\delta(6\phi_0 + 4\phi_0^3\phi_0''') - 3\delta' + \delta'''\phi_0^4 = 0. 
\end{equation} 
Equation (\ref{lin}) is solved using a WKB-type ansatz,  
\begin{equation} 
\delta (\xi) = \epsilon\exp\left\{\frac{\xi^3}{\sqrt{3}}+\dots\right\} . 
\end{equation} 
 
The $O(\xi^0)$ contribution in the exponent turns  
out to be a logarithm, so the full structure is  
\begin{equation}  
\label{delta} 
\delta = \frac{\epsilon}{\xi^2}\exp\left\{ 
\sum_{i=0}^{\infty}\frac{b_i\xi^{3-3i}}{\sqrt{3}}\right\}, 
\end{equation} 
and the coefficients are found to be  
\begin{equation}  
\label{b_i} 
b_0 = 1~, \quad b_2= 0~, \quad b_3 = 32/15~, \quad b_4 = 9\sqrt{3}/5, \dots. 
\end{equation} 
Thus the general form of the solution in the film region  
is  
\begin{equation} 
\phi(\xi) = \phi_0(\xi) + \delta, 
\end{equation} 
with a single free parameter $\epsilon$. An alternative description 
would be an expansion of the form  
\begin{equation}  
\label{expansion} 
\phi(\xi) = \sum_{i=1}^{\infty}\frac{c_i}{\xi^i}, 
\end{equation} 
with $c_1=-1$ and $c_2$ a free parameter. However, the convergence  
of the asymptotic series (\ref{expansion}) turns out to be very bad,  
as perhaps is to be expected from the structure of the WKB solution. 
 
\acknowledgements  
 
We thank Cyprien Gay, Pirouz Kavehpour,
Laurent Limat, Gareth McKinley, Thomas Podgorski and
 David Qu\'er\'e for helpful
conversations. HAS thanks the Harvard MRSEC for partial support of
this research.


\begin{thebibliography}{}

\bibitem[Abraham et al. 2002]{ACM02} 
{\sc Abraham, D. B., Cuerno, R. \& Moro, E.} 2002
Microscopic model for thin film spreading
{\it Phys. Rev. Lett.}  {\bf 88}, 206101(1)-(4). 

\bibitem[Blake \& Ruschak 1979]{BR79} 
{\sc Blake, T. D. \& Ruschak, K. J.} 1979
A maximum speed of wetting.
{\it Nature} {\bf 282}, 489-491. 
  
\bibitem[Chen et al. 2000]{CJV00} 
{\sc Chen, H.-Y., Jasnow, D. \& Vi\~{n}als, J.} 2000
Interface and contact line motion in a two-phase fluid under
shear flow.
{\it Phys. Rev. Lett.} {\bf 85}, 1686-1689.

\bibitem[Chen \& Wada 1989]{CW89} 
{\sc Chen, J.-D. \& Wada, N.} 1989
Wetting dynamics near the edge of a spreading drop.
{\it Phys. Rev. Lett.} {\bf 62}, 3050-3053.

\bibitem[Chen et al. 1995]{CRG95} 
{\sc Chen, Q., Ram\'e, E. \& Garoff, S.} 1995  
The breakdown of asymptotic hydrodynamic models of liquid 
spreading at increasing capillary number.
{\it Phys. Fluids} {\bf 7}, 2631-2639.

\bibitem[Cox 1986]{C86} 
{\sc Cox, R. G.} 1986
The dynamics of the spreading of liquids on a solid surface. 
Part 1. Viscous flow.
{\it J. Fluid Mech.} {\bf 168}, 169-194.

\bibitem[Dussan V. 1979]{D79} 
{\sc Dussan V., E. B.} 1979
On the spreading of liquids on solid surfaces:
static and dynamic contact lines.
{\it Ann. Rev. Fluid Mech.} {\bf 11}, 371-400.

\bibitem[Fermigier \& Jenffer 1991]{FJ91} 
{\sc Fermigier, M. \& Jenffer, P.} 1991
An experimental investigation of the dynamic contact angle  
in liquid-liquid systems.
{\it J. Coll. Int. Sci.} {\bf 146}, 226-241.

\bibitem[de Gennes 1985]{G85} 
{\sc de Gennes, P. G.} 1985 
Wetting: statics and dynamics.
{\it Rev. Mod. Phys.} {\bf 57}, 827-863.

\bibitem[Gorodtsov 1990]{G90} 
{\sc Gorodtsov, V. A.} 1990
Spreading of a film of nonlinearly viscous liquid over a 
horizontal smooth surface.
{\it J. Engrg. Phys.} {\bf 57}, 879-884.

\bibitem[Hervet \& deGennes 1984]{HG84} 
{\sc Hervet, H. \& deGennes, P. G.} 1984
Dynamique du mouillage: films pr\'ecurseurs sur solid `sec'.
{\it C.R. Acad. Sc. Paris, S\'erie II} {\bf 299}, 499-503.

\bibitem[Hocking 1977]{H77} 
{\sc  Hocking, L. M.} 1977
A moving fluid interface. Part 2. The removal of the force 
singularity by a slip flow.
{\it J. Fluid Mech.} {\bf 79}, 209-229.

\bibitem[Hocking 1983]{H83} 
{\sc Hocking, L. M.} 1983
The spreading of a thin drop by gravity and capillarity.
{\it Q. J. Appl. Math.} {\bf 36}, 55-69.

\bibitem[Hocking 1992]{H92} 
{\sc Hocking, L. M.} 1992
Rival contact-angle models and the spreading of drops.
{\it J. Fluid Mech.} {\bf 239}, 671-681.

\bibitem[Hocking 2001]{H01} 
{\sc Hocking, L. M.} 2001 
Meniscus draw-up and draining.  
{\it Euro. J. Appl. Math.} {\bf 12}, 195-208.

\bibitem[Huh \& Mason 1977]{HM77} 
{\sc Huh, C. \& Mason, S. G.} 1977
The steady movement of a liquid meniscus in a capillary tube.
{\it J. Fluid Mech.} {\bf 81}, 401-419.
  
\bibitem[Huh \& Scriven 1971]{HS71} 
{\sc Huh, C. \& Scriven, L. E.} 1971
Hydrodynamic model of steady movement of a solid/liquid/fluid  
contact line. {\it J. Coll. Int. Sci.} {\bf 35}, 85-101.

\bibitem[King 2001]{jking} 
{\sc King, J. R.} 2001
Thin-film flows and high-order degenerate parabolic equations. In
{\it Free Surface Flows}, A. C. King and Y. D. Shikhmurzaev (Eds.),
Kluwer, Dordrecht.

\bibitem[Kistler 1993]{K93} 
{\sc Kistler, S.} 1993
Hydrodynamics of wetting. In 
{\it Wettability}, J. C. Berg (Ed.), Marcel Dekker, New York.

\bibitem[Koplik et al. 1989]{KBW89} 
{\sc Koplik, J., Banavar, J. R. \& Willemsen, J. F.} 1989
Molecular dynamics of fluid flow at solid surfaces 
{\it Phys. Fluids A} {\bf 1}, 781-794.

\bibitem[Leger \& Joanny 1992]{LJ92} 
{\sc Leger, L. \& Joanny, J. F.} 1992
Liquid spreading.
{\it Rep. Prog. Phys.} {\bf 55}, 431-486.

\bibitem[Levich 1962]{L62} 
{\sc Levich, V. G.} 1962 
{\em Physicochemical Hydrodynamics}, 
Prentice-Hall, Englewood Cliffs, N.J.

\bibitem[McKinley \& Ovryn 1998]{mckinley} 
{\sc McKinley, G. H. \& Ovryn, B.} 1998
An interferometric investigation of contact line dynamics in
spreading polymer melts and solutions. In 
{\it Proceedings of the Fourth Microgravity
Fluid Physics and Transport Phenomena Conference}, Cleveland, Ohio.

\bibitem[Marsh et al. 1993]{MGD93} 
{\sc Marsh, J. A., Garoff, 
S. \& Dussan V., E. B.} 1993
Dynamic contact angles and hydrodynamics near a moving contact line.
{\it Phys. Rev. Lett.} {\bf 70}, 2778-2781.
  
\bibitem[Navier 1823]{N23} 
{\sc Navier, C. L.} 1823 (appeared in 1827)
Sur les lois du mouvement des fluides.
{\it Mem. Acad. R. Sci. France} {\bf 6}, 389-440.

\bibitem[Pismen \& Pomeau 2000]{PP00} 
{\sc Pismen, L. M. \& Pomeau, Y.} 2000
Disjoijning potential and spreading of thin layers in the 
diffuse interface model coupled to hydrodynamics. 
{\it Phys. Rev. E} {\bf 62}, 2480-2492.

\bibitem[Podgorski et al. 2001]{PFL01} 
{\sc Podgorski, T., 
Flesselles J. M. \&  Limat, L.} 2001
Corners, cusps, and pearls in running drops.
{\it Phys. Rev. Lett.} {\bf 87}, 036102(1)-(4).

\bibitem[Pomeau 2002]{P02} 
{\sc Pomeau, Y.} 2002
Recent progress in the moving contact line problem: a review
{\it C.R. Mecanique} {\bf 330}, 207-222.
 
\bibitem[Ruijter et al. 1999]{RBC99} 
{\sc Ruijter, M. J., Blake, T. D. \& De Coninck, J.} 1999
Dynamic wetting studied by  molecular modeling simulations 
of droplet spreading.
{\it Langmuir} {\bf 15}, 7836-7847.

\bibitem[Russel et al. 1989]{RSS89} 
{\sc Russel, W. B., Saville, D. A., 
\& Schowalter, W. R.} 1989
{\em Colloidal Suspensions}, p. 148, table 5.3, Cambridge University Press. 

\bibitem[Seppecher 1996]{S96} 
{\sc Seppecher, P.} 1996
Moving contact lines in the Cahn-Hilliard theory.
{\it Int. J. Engng. Sci.} {\bf 34}, 977-992.

\bibitem[Shikhmurzaev 1997]{S97} 
{\sc Shikhmurzaev, Y. D.} 1997
Moving contact lines in liquid/liquid/solid systems.
{\it J. Fluid Mech.} {\bf 334}, 211-249.

\bibitem[Stone et al. 2002]{SLW02} 
{\sc Stone, H. A., Limat, L., Wilson S. K., 
Flesselles J. M. \& Podgorski, T.} 2002
Corner singularity of a contact line moving on a solid substrate.
{\it C. R. Physique} {\bf 3}, 103-110.

\bibitem[Thompson \& Troian 1997]{TT97} 
{\sc Thompson, P. A. \& Troian, S. M.} 1997
A general boundary condition for liquid flow at solid surfaces.
{\it Nature} {\bf 389}, 360-362.

%\bibitem[note]{note}    

\end{thebibliography}
\end{document}